\documentclass[12pt]{article}

\usepackage{array} 
\usepackage{epsfig}
\usepackage{amssymb}
\usepackage{amsmath}                          
\usepackage{graphics,graphpap}
\usepackage{graphicx}
\usepackage{epstopdf}

\renewcommand{\thefootnote}{\fnsymbol{footnote}}

\begin{document}

\begin{titlepage}
\vskip1.5cm
\begin{center}
   {\Large \bf \boldmath Comment on ``Resolving the sign ambiguity in $\Delta \Gamma_s$ with Â
    $B_s \rightarrow D_s K$''}
    \vskip1.3cm {\sc
    Yuehong Xie\footnote{yxie@ph.ed.ac.uk}
\vskip0.3cm    
 {\em  University of Edinburgh, Edinburgh EH9 3JZ, United Kingdom }} \\
\vskip1.5cm 


\vskip2.5cm

{\large\bf Abstract:\\[10pt]} \parbox[t]{\textwidth}{
This is a comment on the recent paper by Soumitra Nandi1 and Ulrich Nierste
``Resolving the sign ambiguity in $\Delta \Gamma_s$ with  $B_s \rightarrow D_s K$'', 
arXiv:0801.0143 [hep-ph].
}

\vfill

\end{center}
\end{titlepage}

\setcounter{footnote}{0}
\renewcommand{\thefootnote}{\arabic{footnote}}

\newpage


In recent paper\cite{Nandi1} Nandi1 and Nierste considered  
the problem of two-fold ambiguity in the quantities extracted from
$B_s \rightarrow J/\psi \phi$: 
$\phi_s \Leftrightarrow \pi - \phi_s$,
$\Delta\Gamma_s \Leftrightarrow -\Delta\Gamma_s$,
$\delta_1 \Rightarrow \pi -\delta_1$ and $\delta_2 \Rightarrow \pi -\delta_2$,
where $\phi_s$ is the $B_s$ mxing phase, $\Delta\Gamma_s = \Gamma_H -\Gamma_L$ is the
decay width difference and $\delta_{1,2}$ are two strong phases. In order to determine 
$sign(\cos\phi_s) = sign (\Delta\Gamma_s)$, one must determine  $sign(\cos\delta_{1,2})$.
This can be done with naive factorisation\cite{Dunietz}. However, the authors argued  
there is no reason to trust naive factorisation and the sign ambiguity in $\Delta \Gamma_s$ and 
$\cos\phi_s$ is unresolved.

The authors proposed to resolve the ambiguity by measuring 
 $$L \equiv b \cos\delta \cos(\phi_s+\gamma)\cos\phi_s$$ 
and
 $$S \equiv b \cos\delta \sin(\phi_s+\gamma)$$
 in $B_s \rightarrow D_s K$, where $b$ is a positive number by definition, 
$\delta$ is a strong phase and the CKM angle $\gamma$ is assumed to be well measured externally.
By assuming that $|\delta|<0.2$, the authors concluded that the value of $S$ ($L$) allows to 
resovle the ambiguity in $sign(\cos\phi_s) = sign (\Delta\Gamma_s)$. 

It should be pointed out that the validity of this conclusion fully depends on the validity of 
the assumption that $|\delta|$ is small. Without this assumption, the value (also sign) of 
$\cos\delta$ is basically unconstrained, therefore  both $S$ and $L$ are allowed to 
take a large range of value, making it very difficult to
resolve the ambiguity in $sign(\cos\phi_s) = sign (\Delta\Gamma_s)$. 

As we know, $\delta \sim 0$ is a result of naive factorisation 
argument\cite{Fleischer}. A natural question to ask is: if we are not prepared to trust 
naive factorisation for $B_s \rightarrow J/\psi \phi$, 
why should we trust it in the case of $B_s \rightarrow D_s K$? 
This means  using $S$ or $L$  alone is insufficient to determine the sign of
$\Delta\Gamma_s$ beyond doubt.

\begin{figure}
\begin{center}
\includegraphics[width=16cm]{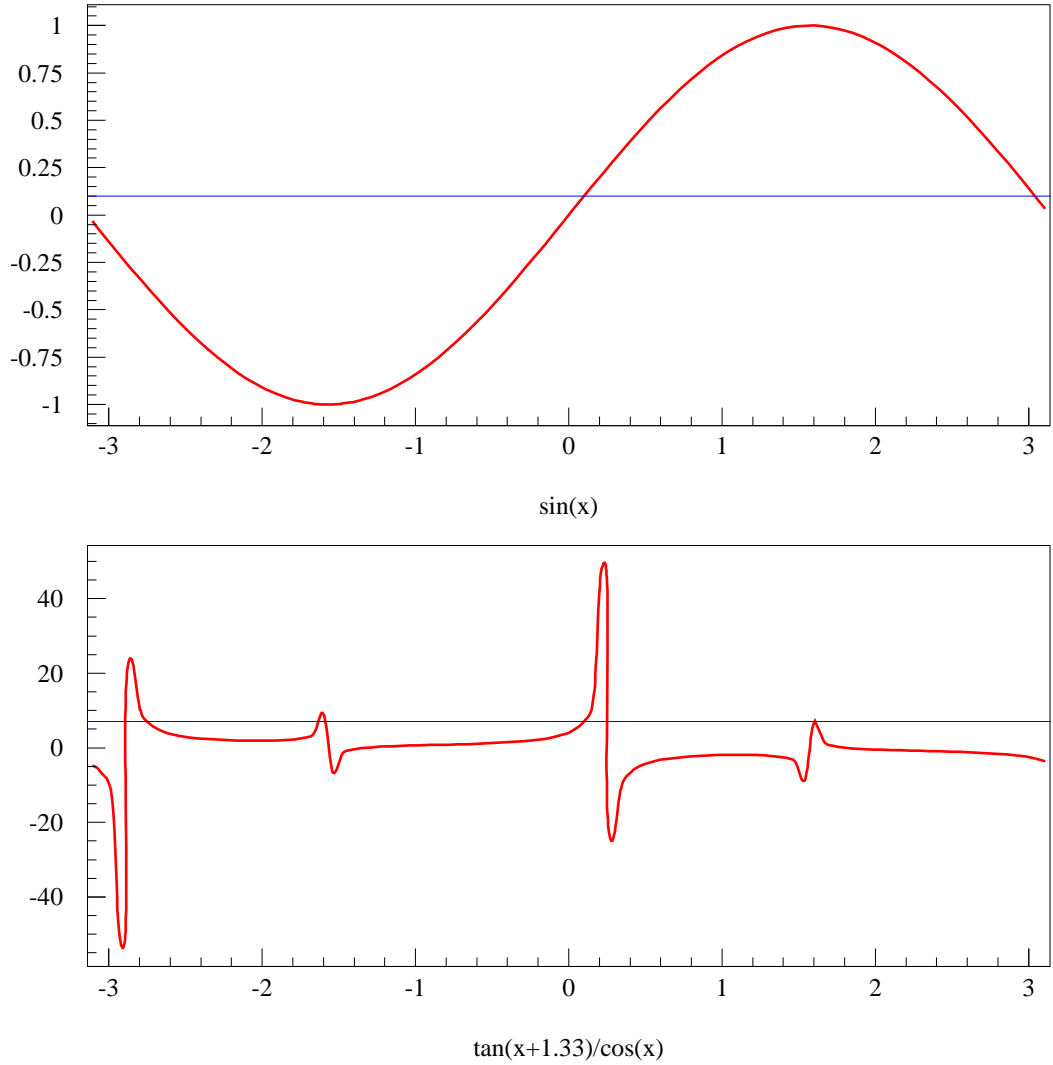}
 \caption{
  Top: red line for $y=\sin(\phi_s)$ from $B_s \rightarrow J/\psi \phi$, 
  blue line for a measurement of $y$ corresponding to $\phi_s = 0.1$; 
  bottom: red line for $y=\tan(\phi_s+\gamma)/\cos\phi_s$ 
  from $B_s \rightarrow D_s K$ with $\gamma = 76 ^{\rm o}$,
  blue line for a  measurement of $y$ corresponding to $\phi_s = 0.1$.
  No easurement error is taken into account.
 }
\label{fig1}
\end{center}
\end{figure}

Fortunately, the authors also suggested it is possible to combine $S$ and $L$ to eliminate
the dependence on $\delta$: 
$$\tan(\phi_s+\gamma) = \frac{S}{L} \cos\phi_s.$$
This allows to determine $\phi_s$ in an unambiguous manner.
As seen in Fig.~\ref{fig1}, there is a two-fold 
ambiguity in $\phi_s$ with $B_s \rightarrow J/\psi \phi$ and an up to eight-fold ambiguity 
in  $\phi_s$ with $B_s \rightarrow D_s K$. If the true value of $\phi_s$ is $0.1$ and 
measurement errors are small enough, then all the discrete ambiguity can be lifted
by combining the two channels.

\end{document}